\documentclass[10pt,twocolumn]{article}          % LaTeX 2e 
\usepackage{latex8}                              % LaTeX 2e 
\usepackage{times}                               % LaTeX 2e 

\usepackage[dvips]{graphicx}
\usepackage{subfigure}
\usepackage[]{rotating}

\begin{document}

\title{ Design and Implementation of MPICH2 over InfiniBand with RDMA
Support\thanks{This research is supported in part by Department of
Energy's Grant \#DE-FC02-01ER25506, a grant from Sandia National
Laboratories, a grant from Intel Corporation, National Science
Foundation's grants \#CCR-0204429 and \#CCR-0311542, and by the
Mathematical, Information, and Computational Sciences Division
subprogram of the Office of Advanced Scientific Computing Research,
Office of Science, U.S. Department of Energy, under Contract
W-31-109-ENG-38.}  }

\author{ \hspace*{3ex}Jiuxing Liu{\textsuperscript{\dag}} 
\and Weihang Jiang{\textsuperscript{\dag}}
\and Pete Wyckoff{\textsuperscript{\ddag}}
\and Dhabaleswar K.~Panda{\textsuperscript{\dag}}\hspace*{3ex}
\and David Ashton{\textsuperscript{**}}
\and Darius Buntinas{\textsuperscript{**}}
\and William Gropp{\textsuperscript{**}}
\and Brian Toonen{\textsuperscript{**}} \and \\
{\textsuperscript{\dag}}Computer and Information Science,
The Ohio State University,
Columbus, OH\ 43210 \\
\{liuj, jiangw, panda\}@cis.ohio-state.edu \and \\
{\textsuperscript{\ddag}}Ohio Supercomputer Center,
1224 Kinnear Road,
Columbus, OH\  43212\\
pw@osc.edu \and \\
{\textsuperscript{**}}Mathematics and Computer Science Division,
Argonne National Laboratory,
Argonne, IL\ 60439\\
\{ashton, buntinas, gropp, toonen\}@mcs.anl.gov
}

\maketitle

\vspace*{-5.0ex}
\begin{abstract}
{ 

For several years, MPI has been the de facto standard for writing
parallel applications.  One of the most popular MPI implementations is
MPICH. Its successor, MPICH2, features a completely new design that
provides more performance and flexibility. To ensure portability, it
has a hierarchical structure based on which porting can be done
at different levels.

In this paper, we present our experiences designing and implementing
MPICH2 over InfiniBand.  Because of its high performance and open
standard, InfiniBand is gaining popularity in the area of
high-performance computing.  Our study focuses on optimizing the
performance of MPI-1 functions in MPICH2. One of our objectives is to
exploit Remote Direct Memory Access (RDMA) in Infiniband to achieve
high performance. We have based our design on the RDMA Channel
interface provided by MPICH2, which encapsulates
architecture-dependent communication functionalities into a very small
set of functions.

Starting with a basic design, we apply different optimizations and
also propose a zero-copy-based design. We characterize the impact of
our optimizations and designs using microbenchmarks.  We have also
performed an application-level evaluation using the NAS Parallel
Benchmarks. Our optimized MPICH2 implementation achieves 7.6~$\mu$s
latency and 857~MB/s bandwidth, which are close to the raw performance
of the underlying InfiniBand layer.  Our study shows that the RDMA
Channel interface in MPICH2 provides a simple, yet powerful,
abstraction that enables implementations with high performance by
exploiting RDMA operations in InfiniBand.  To the best of our
knowledge, this is the first high-performance design and
implementation of MPICH2 on InfiniBand using RDMA support.

}
\end{abstract}

\section{Introduction}
\label{sec:intro}

During the past ten years, the research and industry communities have
proposed and implemented user-level communication systems to address
some of the problems associated with traditional networking protocols.
The Virtual Interface Architecture (VIA)~\cite{via_dunning98} was
proposed to standardize these efforts.  More recently, the InfiniBand
Architecture~\cite{IB-SPEC} has been introduced, which combines storage
I/O with interprocess communication.
                                                                                
In addition to send and receive operations, InfiniBand architecture
supports Remote Direct Memory Access (RDMA). RDMA operations enable
direct access to the address space of a remote process. These operations introduce
new opportunities and challenges in designing communication protocols. 
                                                                                
In the area of high-performance computing, MPI~\cite{mpi-ref} has been
the de facto standard for writing parallel applications. After the
original MPI standard (MPI-1), an enhanced standard
(MPI-2)~\cite{MPI2} was introduced, which includes features such
as dynamic process management, one-sided communication, and I/O.
MPICH~\cite{MPICH} is one of the most popular MPI-1 implementations.
Recently, work has begun to create MPICH2~\cite{MPICH2}, which aims
to support both MPI-1 and MPI-2 standards. It features a completely
new design that provides better performance and flexibility than its
predecessor. MPICH2 is also very portable and provides mechanisms
that make it easy to retarget MPICH2 to new communication architectures
such as InfiniBand.

In this paper, we present our experiences in designing and
implementing MPICH2 over InfiniBand using RDMA operations. Although
MPICH2 supports both MPI-1 and MPI-2, our study focuses on optimizing
the performance of MPI-1 functions.  We have based our design on the
RDMA Channel interface provided by MPICH2, which encapsulates
architecture-dependent communication functionalities in a small set of
functions.  Despite its simple interface, we have shown that the RDMA
Channel does not prevent one from achieving high performance. In our
testbed, our MPICH2 implementation achieves 7.6~$\mu$s latency and
857~MB/s peak bandwidth, which are quite close to the raw performance
of the InfiniBand platform.  We have also evaluated our designs using
the NAS Parallel Benchmarks~\cite{NAS_BENCH}. Overall, we have
demonstrated that the RDMA Channel interface is a simple, yet
powerful, abstraction that makes it possible to design
high-performance MPICH2 implementations with less development effort.

In our design, communication between processes is done exclusively
using RDMA operations.
Our design starts with an emulation of a shared-memory-based
implementation.  Then we introduce various optimization techniques to
improve its performance.  To evaluate the impact of each optimization,
we use latency and bandwidth microbenchmarks.  We also propose a
zero-copy design for large messages.  Our results show that with {\em
piggybacking} and {\em zero-copy} optimizations for large messages,
our design achieves very good performance.

The remainder of the paper is organized as follows.  In Section 2, we
provide an introduction to InfiniBand and its RDMA operations.  In
Section 3, we present an overview of MPICH2, its implementation
structure, and the RDMA Channel interface. In Sections~4 and 5, we
describe our designs and implementations. In Section 6, we compare our
RDMA Channel-based design with another design based on a more
complicated interface called CH3. In Section 7, we present an
application level performance evaluation. In Section 8, we describe
related work. In Section 9, we draw conclusions and briefly mention
some future research directions.

\section{InfiniBand Overview}
\label{sec:iba}

The InfiniBand Architecture (IBA)~\cite{IB-SPEC} defines a switched
network fabric for interconnecting processing nodes and I/O nodes.  It
provides a communication and management infrastructure for
interprocessor communication and I/O.
In an InfiniBand network, processing nodes and I/O nodes are connected
to the fabric by channel adapters (CAs). Channel adapters usually have
programmable DMA engines with protection features.
There are two kinds of channel adapters: host channel adapter (HCA)
and target channel adapter (TCA). HCAs sit on processing nodes.

The InfiniBand communication stack consists of different layers. The
interface presented by channel adapters to consumers belongs to the
transport layer.  A queue-based model is used in this interface. A
queue pair in InfiniBand Architecture consists of two queues: a send
queue and a receive queue. The send queue holds instructions to
transmit data, and the receive queue holds instructions that describe
where received data is to be placed. Communication operations are
described in work queue requests (WQRs), or descriptors, and are
submitted to the work queue.
The completion of WQRs is reported through completion queues
(CQs). Once a work queue element is finished, a completion queue entry
is placed in the associated completion queue. Applications can check
the completion queue to see whether any work queue request has been
finished. InfiniBand also supports different classes of transport
service. In this paper, we focus on the reliable connection (RC)
service.

\subsection{RDMA Operations in InfiniBand Architecture}

InfiniBand Architecture supports both channel and memory semantics. In
channel semantics, send/receive operations are used for
communication. To receive a message, the programmer posts a receive
descriptor that describes where the message should be put at the
receiver side. At the sender side, the programmer initiates the send
operation by posting a send descriptor.

In memory semantics, InfiniBand supports remote direct memory access
(RDMA) operations, including RDMA write and RDMA read.  RDMA
operations are one sided and do not incur software overhead at the
remote side.  In these operations, the sender (initiator) starts RDMA
by posting RDMA descriptors. A descriptor contains both the local data
source addresses (multiple data segments can be specified at the
source) and the remote data destination address.  At the sender side,
the completion of an RDMA operation can be reported through CQs.  The
operation is transparent to the software layer at the receiver side.

Since RDMA operations enable a process to access the address space of
another process directly, they have raised some security concerns.  In
InfiniBand architecture, a key-based mechanism is used.  A memory
buffer must first be registered before it can be used for
communication.  Among other things, the registration generates a
remote key.  This remote key must be presented when the sender
initiates an RDMA operation to access the buffer.

Compared with send/receive operations, RDMA operations have several
advantages.  First, RDMA operations themselves are generally faster
than send/receive operations because they are simpler at the hardware
level.  Second, they do not involve managing and posting descriptors
at the receiver side, which would incur additional overheads and
reduce the communication performance.

\section{MPICH2 Overview}
\label{sec:mpich2}

MPICH~\cite{MPICH} is developed at Argonne National Laboratory. It is
one of the most popular MPI implementations. The original MPICH
provides support for the MPI-1 standard. As a successor of MPICH,
MPICH2~\cite{MPICH2} aims to support not only the MPI-1 standard but
also functionalities such as dynamic process management, one-sided
communication, and MPI I/O, which are specified in the MPI-2
standard. However, MPICH2 is not merely MPICH with MPI-2
extensions. It is based on a completely new design, aiming to provide
more performance, flexibility, and portability than the original MPICH.
One of the notable features in the implementation of MPICH2 is that it
can take advantage of RDMA operations if they are provided by the
underlying interconnect. These operations can be used not only to
support MPI-2 one-sided communication but also to implement normal
MPI-1 communication.  Although MPICH2 is still under development, beta
versions are already available for developers. In the current version,
all MPI-1 functions have been implemented. MPI-2 functions are not
completely supported yet. In this paper, we mainly focus on the MPI-1
part of MPICH2.

\subsection{MPICH2 Implementation Structure}

One of the objectives in MPICH2 design is portability. To facilitate
porting MPICH2 from one platform to another, MPICH2 uses ADI3 (the
third generation of the Abstract Device Interface) to provide a
portability layer.  ADI3 is a full-featured abstract device interface
and has many functions, so it is not a trivial task to implement all
of them. To reduce the porting effort, MPICH2 introduces the CH3
interface.  CH3 is a layer that implements the ADI3 functions and
provides an interface consisting of only a dozen functions.  A
``channel'' implements the CH3 interface.  Channels exist for
different communication architectures such as TCP sockets and, SHMEM.
Because only a dozen functions are associated with each channel
interface, it is easier to implement a channel than the ADI3 device.

To take advantage of architectures with globally shared memory or RDMA
capabilities and to further reduce the porting overhead, MPICH2
introduces the RDMA Channel which implements the CH3 interface. The
RDMA Channel interface only contains five functions. We will discuss
the details of the RDMA Channel interface in the next subsection.

\begin{figure}[htbp]
\includegraphics[width=\columnwidth]{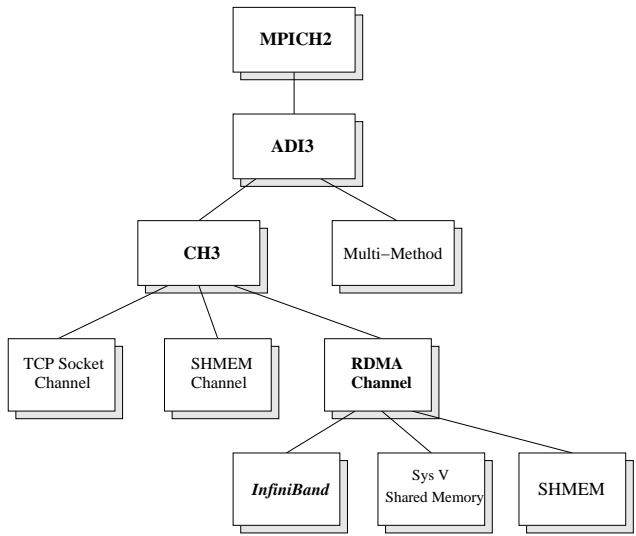}
\caption{MPICH2 Implementation Structure}
\label{fig:mpich2}
\end{figure}

The hierarchical structure of MPICH2 , as shown in
Figure~\ref{fig:mpich2}, gives much flexibility to implementors. The
three interfaces (ADI3, CH3, and the RDMA Channel interface) provide
different trade-offs between communication performance and ease of
porting.

\subsection{MPICH2 RDMA Channel Interface}

The MPICH2 RDMA Channel interface is designed specifically for
architectures with globally shared memory or RDMA capabilities. It
contains five functions, among which only two are central to
communication. (Other functions deal with process management,
initialization, and finalization.)  These two functions are called
{\em put (write)} and {\em get (read)}.

Both {\em put} and {\em get} functions accept a connection structure
and a list of buffers as parameters. They return to the caller the
number of bytes that have been successfully put or gotten. If the
number of bytes completed is less than the total length of buffers,
the caller will retry the same {\em get} or {\em put} operation later.

\begin{figure}[htbp]
\center
\includegraphics[width=0.9\columnwidth]{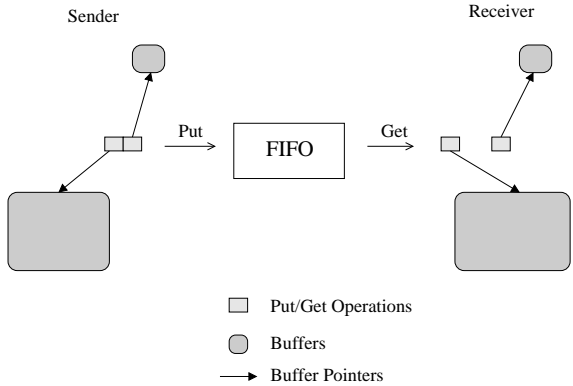}
\caption{Put and Get Operations}
\label{fig:put_get}
\end{figure}

Figure~\ref{fig:put_get} illustrates the semantics of {\em put} and
{\em get}.  Logically, a pipe is shared between the sender and the
receiver.  The {\em put} operation writes to the pipe, and the {\em
get} operation reads from it. The data in the pipe is consumed in FIFO
order.  Both operations are nonblocking in the sense that they return
immediately with the number of bytes completed, instead of waiting for
the entire operation to finish.  We note that {\em put} and {\em get}
are different from RDMA write and RDMA read in InfiniBand. While RDMA
operations in InfiniBand are one sided, {\em put} and {\em get} in the
RDMA Channel interface are essentially two-sided operations.

\begin{figure}[htbp]
\center
\includegraphics[width=0.8\columnwidth]{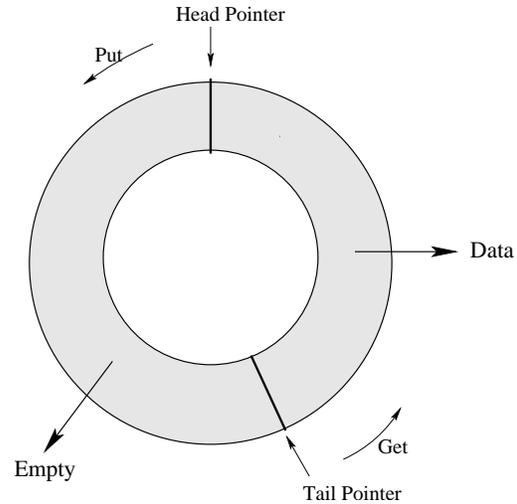}
\caption{Put and Get Implementation with Globally Shared Memory}
\label{fig:shmem}
\end{figure}

{\em Put} and {\em get} operations can be implemented on architectures
with globally shared memory in a straightforward
manner. Figure~\ref{fig:shmem} shows one example. In this
implementation, a shared buffer (organized logically as a ring) is
placed in shared memory, together with a head pointer and a tail
pointer.  The {\em put} operation copies the user buffer to the shared
buffer and adjusts the head pointer. The {\em get} operation involves
reading from the shared buffer and adjusting the tail pointer.  In the
case of buffer overflow or underflow (detected by comparing head and
tail pointers), the operations return immediately, and the caller will
retry them.

Working at the RDMA Channel interface level is better than writing a
new CH3 or ADI3 implementation for many reasons:
\begin{enumerate}
\item Improvements done at this level can affect all
shared-memory-like transports such as globally shared memory, RDMA
over IP, Quadrics, and Myrinet.

\item Other protocols on InfiniBand need efficient processing,
including one-sided communication in MPI-2, DSM systems, and parallel
file systems. The RDMA Channel interface can potentially be used also
for them.

\item Designing proper interfaces to similar systems improves
performance and portability in general.
\end{enumerate}

In collaboration, the OSU and ANL teams 
are also currently working together to design an improved interface that
can benefit communication systems in general.

\section{Designing and Optimizing MPICH2 over InfiniBand}
\label{sec:design}

In this section, we present several different designs of MPICH2 over
InfiniBand based on the RDMA Channel interface.
We first start with a basic design that resembles the scheme described
in Figure~\ref{fig:shmem}. Then we apply various optimization
techniques to improve its performance.  In this section, the designs
are evaluated by using microbenchmarks such as latency and bandwidth.
We show that by taking advantage of RDMA operations in InfiniBand, we
can achieve not only low latency for small messages but also high
bandwidth for large messages using the RDMA Channel interface.  In
Section 5, we present a zero-copy design.

\subsection{Experimental Testbed}
Our experimental testbed consists of a cluster system with 8
SuperMicro SUPER P4DL6 nodes. Each node has dual Intel Xeon 2.40 GHz
processors with a 512K L2 cache and a 400 MHz front side bus.  The
machines are connected by Mellanox InfiniHost MT23108 DualPort 4X HCA
adapter through an InfiniScale MT43132 Eight 4x Port InfiniBand
Switch. The HCA adapters work under the PCI-X 64-bit 133MHz
interfaces.
We used the Linux Red Hat 7.2 operating system with 2.4.7 kernel.
The compilers we used were GNU GCC 2.96 and GNU FORTRAN 0.5.26.

\subsection{Basic Design}

In Figure~\ref{fig:shmem}, we illustrated how the RDMA Channel
interface can be implemented on shared-memory architectures. In a
cluster connected by InfiniBand, however, there is no physically
shared memory.  In our basic design, we use RDMA write operations
provided by InfiniBand to emulate this scheme.

We put the shared-memory buffer in the receiver's main memory. This
memory is registered and exported to the sender. Therefore, it is
accessible to the sender through RDMA operations. To avoid the
relatively high cost of registering user buffers for sending every
message, we also use a preregistered buffer at the sender that is the
same size as the shared buffer at the receiver. User data is first
copied into this buffer and then sent out.  Head and tail pointers
also need to be shared between the sender and the receiver. Since they
are used frequently at both sides, we use replication to prevent
polling through the network. For the tail pointer, a master copy is
kept at the receiver, and a replica is kept at the sender. For the
head pointer, a master copy is kept at the sender, and a replica is
kept at the receiver. For each direction of every connection, the
associated ``shared'' memory buffer, head and tail pointers are
registered during initialization, and their addresses and remote keys
are exchanged.

At the sender, the {\em put} operation is implemented as follows:
\begin{enumerate}
	\item Use local copies of head and tail pointers to decide
		how much empty space is available.
	\item Copy user buffer to the preregistered buffer.
	\item Use RDMA write operation to write the data to the
		buffer at the receiver side.
	\item Adjust the head pointer based on the amount of data
		written.
	\item Use another RDMA write to update the remote copy
		of head pointer.
	\item Return the number of bytes written.
\end{enumerate}

At the receiver, the {\em get} operation is implemented in the following way:
\begin{enumerate}
	\item Check local copies of head and tail pointers to 
		see whether there is new data available.
	\item Copy the data from the shared memory buffer to
		user buffer.
	\item Adjust the tail pointer based on the amount of data
		that has been copied.
	\item Use an RDMA write to update the remote copy of tail
		pointer.
	\item Return the number of bytes successfully read.
\end{enumerate}

We note that copies of head and tail pointers are not always
consistent. For example, after a sender adjusts its head pointer, it
uses RDMA write to update the remote copy at the receiver. Therefore,
the head pointer at the receiver is not up to date until the RDMA
write finishes. However, this inconsistency does not affect the
correctness of the scheme; it merely prevents the receiver from
reading new data temporarily. Similarly, inconsistency of tail pointer
may prevent the sender from writing to the shared buffer. But
eventually the pointers will become up to date, and the sender or the
receiver will be able to make progress.

\subsubsection{Performance of the Basic Design}

We use latency and bandwidth tests to evaluate the performance of our
basic design. The latency test is conducted in a ping-pong fashion,
and the results are derived from round-trip time. In the bandwidth
test, a sender keeps sending back-to-back messages to the receiver
until it has reached a predefined window size {\em W}.  Then it waits
for these messages to finish and send out another {\em W} messages.
The results are derived from the total test time and the number of
bytes sent.

\begin{figure}[htbp]
\includegraphics[width=\columnwidth]{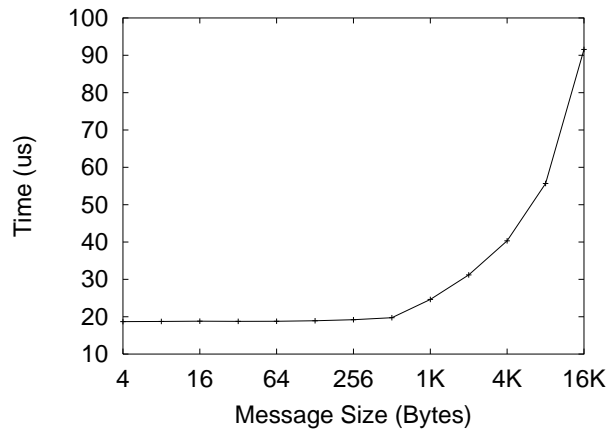}
\caption{MPI Latency for Basic Design}
\label{fig:basic_lat}
\end{figure}

\begin{figure}[htbp]
\includegraphics[width=\columnwidth]{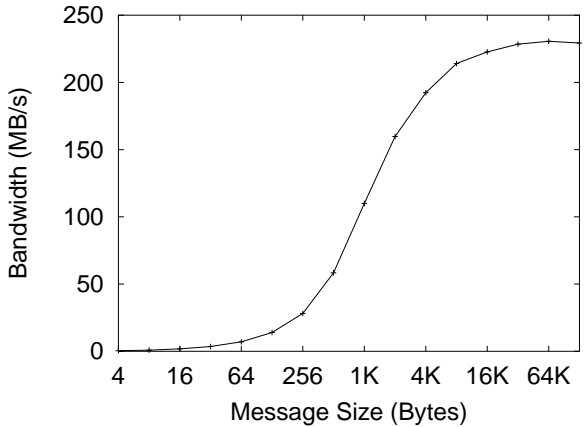}
\caption{MPI Bandwidth for Basic Design}
\label{fig:basic_bw}
\end{figure}

Figures~\ref{fig:basic_lat} and \ref{fig:basic_bw} show the results.
Our basic design achieves a latency of 18.6~$\mu$s for small messages
and a bandwidth of 230~MB/s for large messages.  (Unless stated
otherwise, the unit MB in this paper is an abbreviation for $10{}^{6}$
bytes, not $2^{20}$ bytes.)  These numbers are much worse than the raw
performance numbers achievable by the underlying InfiniBand layer
(5.9~$\mu$s latency and 870~MB/s bandwidth).

A careful look at the basic design reveals many inefficiencies.  For
example, a matching pair of send and receive operations in MPI require
three RDMA write operations to take place: one for transfer of data,
and two for updating head and tail pointers. These not only increase
latency and host overhead but also generate unnecessary network
traffic.

For large messages, the basic scheme leads to two extra memory
copies. The first one is from user buffer to the preregistered buffer
at the sender side. The second one is from the shared buffer to user
buffer at the receiver side. These memory copies consume resources
such as memory bandwidth and CPU cycles. To make matters worse, in the
basic design the memory copies and communication operations are
serialized. For example, a sender first copies the whole message (or
part of the message if it cannot fit itself in the empty space of the
preregistered buffer). Then it initiates RDMA write to transfer the
data. This serialization of copying and RDMA write greatly reduces the
bandwidth for large messages.

\subsection{Optimization with Piggybacking Pointer Updates}

Our first optimization targeted to avoid separate
head and tail pointer updates whenever
possible. The technique we used is
piggybacking, which combines pointer updates
with data transfer.

At the sender side, we combine data and the new value of head pointer
into a single message.  To help the receiver detect the arrival of the
message, we attach the size with the message and put two flags at the
beginning and the end of the message. The receiver detects arrival of
the new message by polling on the flags. To avoid possible situations
where the buffer content happens to have the same value as the flag,
we divide the shared buffer into fixed-sized chunks.  Each message
uses a different chunk. In this way, the situations can be handled by
using two polling flags, or ``bottom fill.''  Similar techniques have
been used in ~\cite{MVAPICH_RDMA,Rinku-ipdps03}.

At the receiver side, instead of using RDMA write to update the remote
tail pointer each time data has been read, we delay the updates until
the free space in the shared buffer drops below a certain
threshold. If messages are sent from the receiver to the sender, the
pointer value is attached with the message, and no extra message is
used to transfer pointer updates. If no messages are sent from the
receiver to the sender, eventually we will explicitly send the updates
by using an extra message. The sender updates its pointer after
receiving this message.  Even in this case, however, the traffic can be
reduced because several consecutive updates of the tail pointer can be
sent by using only one message.

\begin{figure}[htbp]
\includegraphics[width=\columnwidth]{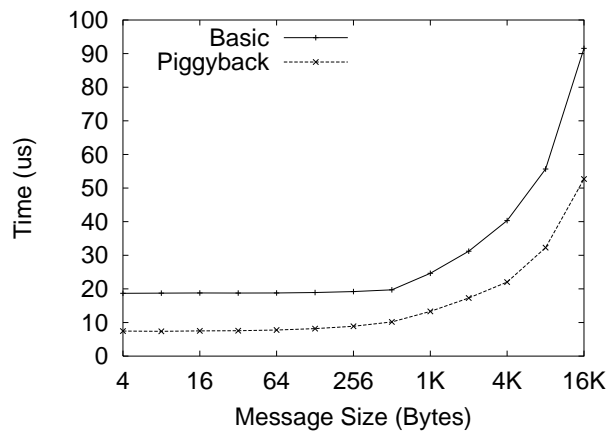}
\caption{MPI Small-Message Latency with Piggybacking}
\label{fig:piggyback_lat}
\end{figure}

\begin{figure}[htbp]
\includegraphics[width=\columnwidth]{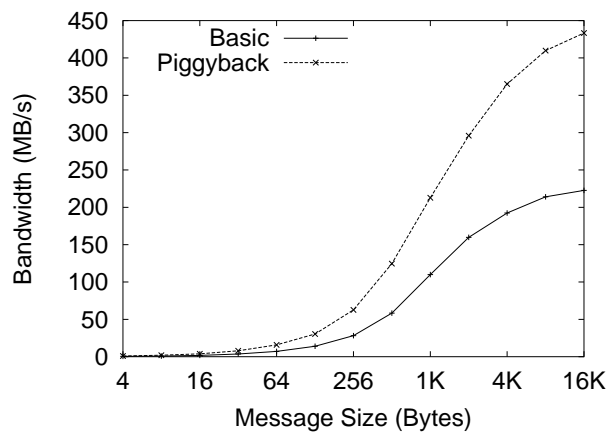}
\caption{MPI Small-Message Bandwidth with Piggybacking}
\label{fig:piggyback_bw}
\end{figure}

The use of piggybacking and delayed pointer updates can greatly
improve the performance of small message. From
Figure~\ref{fig:piggyback_lat} we see that the latency is reduced from
18.6~$\mu$s to 7.4~$\mu$s.  Figure~\ref{fig:piggyback_bw} shows that
the optimization also improves bandwidth for small messages.

\subsection{Optimization with Pipelining of Large Messages}

As we have discussed, our basic design suffers from serialization of
memory copies and RDMA writes. A better solution is to use pipelining
to overlap memory copies with RDMA write operations.

In our piggybacking optimization, we divide the shared-memory buffer
into small chunks. When sending and receiving large messages, we need
to use more than one such chunks. At the sender side, instead of
starting RDMA writes after copying all the chunks, we initiate the
RDMA transfer immediately after copying each chunk. In this way, the
RDMA operation can be overlapped with the copying of the next
chunk. Similarly, at the receiver side we start copying from the
shared buffer to the user buffer immediately after a chunk is
received. In this way, the receive RDMA operations can be overlapped
with the copying.

\begin{figure}[htbp]
\includegraphics[width=\columnwidth]{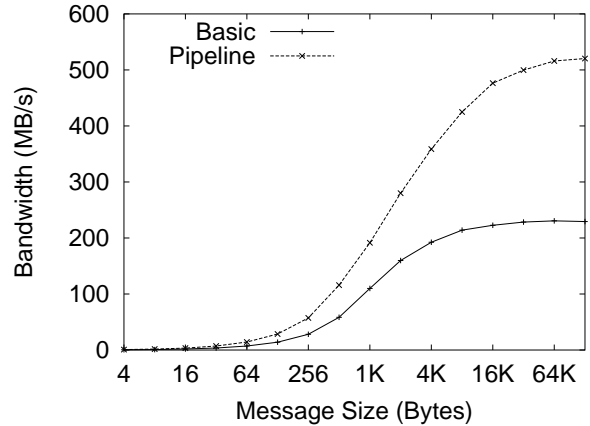}
\caption{MPI Bandwidth with Pipelining}
\label{fig:pipe_bw}
\end{figure}

Figure~\ref{fig:pipe_bw} compares the bandwidth of the pipelining
scheme with the basic scheme. (Piggybacking is also used in the
pipelining scheme.) We can see that pipelining combined with
piggybacking has greatly improved MPI bandwidth. The peak bandwidth
has been increased from 230~MB/s to over 500~MB/s.  This result is
still not satisfying, however, because InfiniBand is able to deliver
bandwidth up to 870~MB/s.

To investigate the performance bottleneck, we have conducted memory
copy tests in our testbed.  We have found that memory copy bandwidth
is less than 800~MB/s for large messages.  In our MPI bandwidth tests,
with RDMA write operations and memory copies both using the memory
bus, the bandwidth achievable at the application level is even less.
Therefore, the memory bus clearly becomes a performance bottleneck for
large messages because of the extra memory copies.

\begin{figure}[htbp]
\includegraphics[width=\columnwidth]{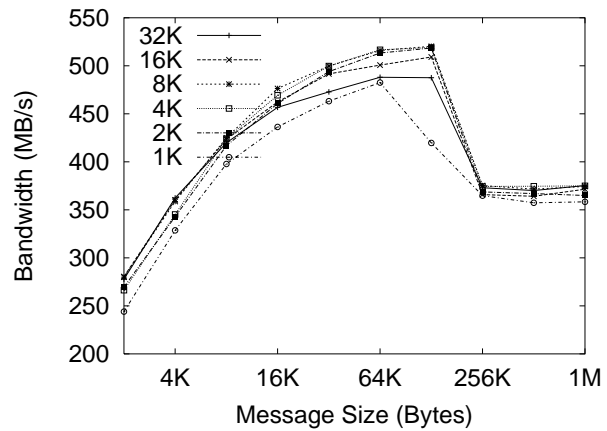}
\caption{MPI Bandwidth with Pipelining (Different Chunk Sizes)}
\label{fig:pipeline_bufsize}
\end{figure}

In the pipelining optimization, it is important that we balance each
stage of the pipeline so that we can get maximum throughput.  One
parameter we can change to balance pipeline stages is the chunk size,
or how much data we copy each time for a large message.
Figure~\ref{fig:pipeline_bufsize} shows MPI bandwidth for different
chunk sizes for the pipelining optimization. We observe that MPI does
not give good performance when the chunk size is either too small (1K
Bytes) or too large (32K Bytes).  MPI performs comparably for chunk
sizes of 2K to 16K Bytes. In all remaining tests, we have chosen a
chunk size of 16K Bytes.

\section{Zero-Copy Design}

As we have discussed in the preceding section, it is desirable to
avoid memory copies for large messages.  In this section, we describe
a zero-copy design for large messages based on the RDMA Channel
interface.

In our new design, small messages are still transferred by using RDMA
write, similar to the piggybacking scheme.  For large messages,
however, RDMA read, instead of RDMA write, is used for data
transfer. The basic idea of our zero-copy design is to let the
receiver ``pull'' the data directly from the sender using RDMA read.

For each connection, shared buffers are still used for transferring
small messages.  However, the data for large messages is not
transferred through the shared buffer.  At the sender, when the {\em
put} function is called, it checks the user buffer and decides whether
to use zero-copy or not, based on the buffer size.  If zero-copy is
not used, the message is sent through the shared buffer as discussed
before.  If zero-copy is used, the function registers the user buffer,
constructs a special packet that contains information about the user
buffer such as address, size, and remote key, then sends the special
packet by using RDMA write through the shared buffer.  The {\em put}
function returns a value of 0, at this stage, because no data has been
transferred yet.  Subsequent calls to {\em put} also return 0 until
all of the data has been transferred, and the operation has completed.
Once the operation has completed, {\em put} will return the number of
bytes transferred.

\begin{figure}[htbp]
\center
\includegraphics[width=0.8\columnwidth]{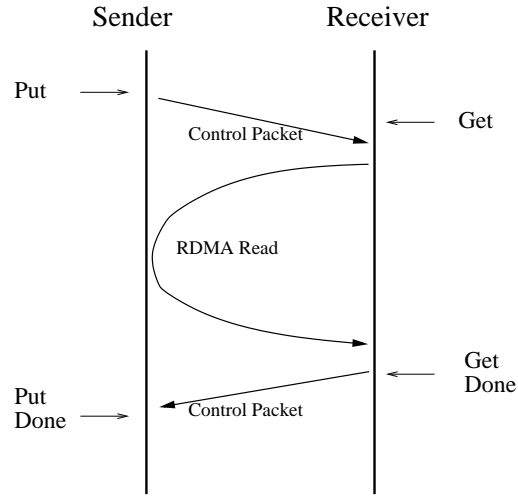}
\caption{Zero-Copy Design}
\label{fig:zcopy}
\end{figure}

When the packet arrives at the other side and the {\em get} function
is called, the receiver checks the shared buffer and processes all the
packets in order.  If a packet is a data packet, the data is copied to
the user buffer.  If it is a special packet, the user buffer is
registered, and an RDMA read operation is issued to fetch the data
from the remote side directly to the user buffer.  After initiating
the RDMA read, the {\em get} function returns with a value of 0,
because the operation is still in progress.  When the RDMA read is
finished, calling the {\em get} function leads to an acknowledgment
packet being sent to the sender.  The {\em get} function then returns
the number of bytes successfully transferred.  When the acknowledgment
packet is received at the sender side, the sender deregisters the user
buffer, completing the operation, and the next call to the {\em put}
function will return the number of bytes transferred.  The zero-copy
process is illustrated in Figure~\ref{fig:zcopy}.

In the current InfiniBand implementation, memory registration and
deregistration are expensive operations. To reduce the number of
registrations and deregistrations, we have implemented a registration
cache~\cite{PINDOWN-CACHE}. The basic idea is to delay the
deregistration of user buffers and put them into a cache. If the same
buffer is reused later, its registration information can be fetched
directly from the cache instead of going through the expensive
registration process. Deregistration happens only when there are too
many registered user buffers.

We note that the effectiveness of registration cache depends on
buffer reuse patterns of applications. If applications rarely reuse
buffers for communication, registration overhead cannot be avoided most
of the time. Fortunately, our previous study with the NAS Parallel 
Benchmarks~\cite{MVAPICH_SC} has demonstrated that buffer reuse rates
are very high in these applications.

\begin{figure}[htbp]
\includegraphics[width=\columnwidth]{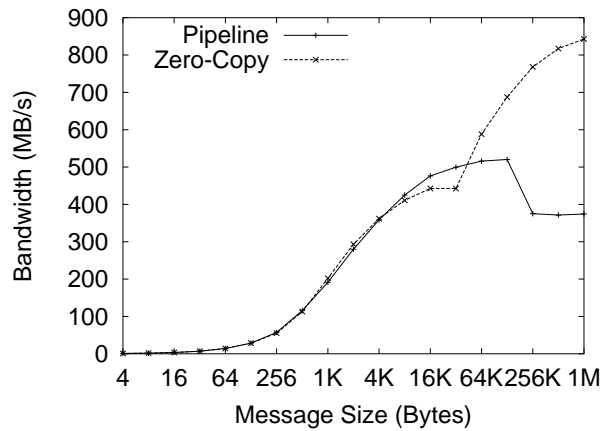}
\caption{MPI Bandwidth with Zero-Copy and Pipelining}
\label{fig:zcopy_bw}
\end{figure}

We compare the bandwidth of the pipelining design and the zero-copy
design in Figure~\ref{fig:zcopy_bw}. We observe that zero-copy greatly
improves the bandwidth for large messages. We achieve a peak bandwidth
of 857~MB/s, which is quite close to the peak bandwidth at the
InfiniBand level (870~MB/s). We also see that as a result of cache
effect, bandwidth for large messages drops for the pipelining
design. Because of the extra overhead in the implementation, the
zero-copy design slightly increases the latency for small messages,
which is now around 7.6~$\mu$s.

Our zero-copy implementation uses RDMA read operations, which let the
receiver ``pull'' data from the sender.  An alternative is to use RDMA
write operations and let the sender ``push'' data to the receiver.
Before the sender can push the data, the receiver has to use special
packets to advertise availability of new receive buffers.  Therefore,
this method can be very efficient if the {\em get} operations are
called {\em before} the corresponding {\em put} operations.  In the
current MPICH2 implementation, however, the layers above the RDMA
Channel interface are implemented in such a way that {\em get} is
always called after {\em put} for large messages.  Therefore, we have
chosen an RDMA read-based implementation instead of RDMA write.

\section{Comparing CH3 and RDMA Channel Interface Designs}

The RDMA Channel interface in MPICH2 provides a simple way to
implement MPICH2 in many communication architectures.  In the
preceding section, we showed that this interface does not prevent one
from achieving good performance. Nor does it prevent zero-copy
implementation for large messages. Our results showed that with
various optimizations, we can achieve a latency of 7.6~$\mu$s and a
peak bandwidth of 857~MB/s.

The CH3 interface is more complicated than the RDMA Channel
interface. Therefore, porting it requires more effort. However, since
CH3 provides more flexibility, it is possible to achieve better
performance at this level.

\begin{figure}[htbp]
\center
\includegraphics[width=0.8\columnwidth]{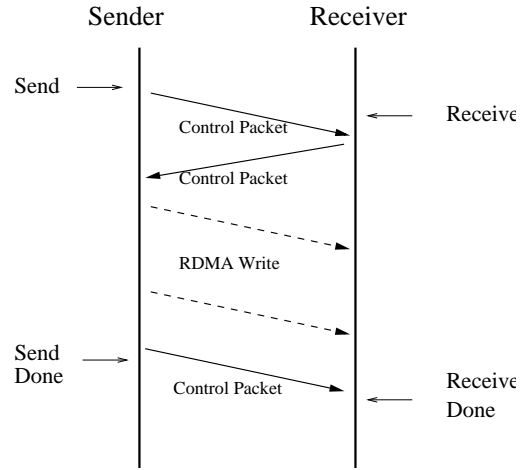}
\caption{CH3 Zero-Copy with RDMA Write}
\label{fig:rdmawrite}
\end{figure}

To study the impact of different interfaces on MPICH2 performance, we
have also done a CH3-level implementation. This implementation uses
RDMA write operations for transferring large messages, as shown in
Figure~\ref{fig:rdmawrite}. Before transferring the message, a
handshake happens between the sender and the receiver. User buffer at
the receiver is registered and its information is sent to the sender
through the handshake. The sender then uses RDMA write to transfer the
data. A registration cache is also used in this implementation.

\begin{figure}[htbp]
\includegraphics[width=\columnwidth]{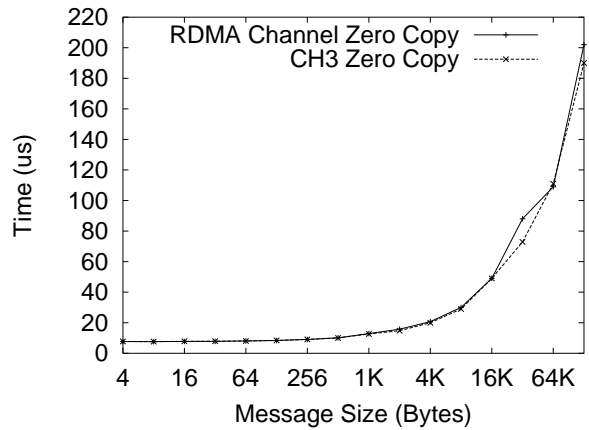}
\caption{MPI Latency for CH3 Design and RDMA Channel Interface Design}
\label{fig:rdmawrite_lat}
\end{figure}

\begin{figure}[htbp]
\includegraphics[width=\columnwidth]{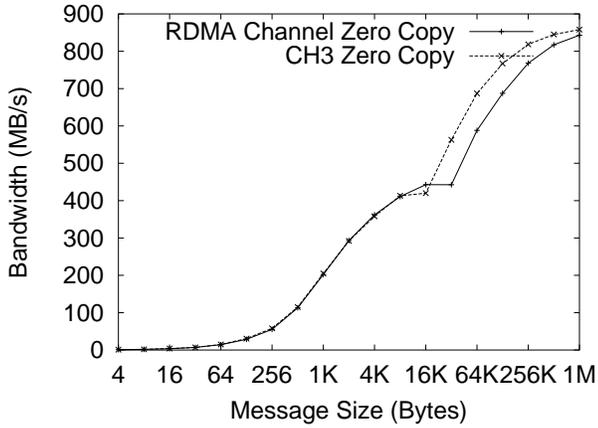}
\caption{MPI Bandwidth for CH3 Design and RDMA Channel Interface Design}
\label{fig:rdmawrite_bw}
\end{figure}

Figures~\ref{fig:rdmawrite_lat} and \ref{fig:rdmawrite_bw} compare
this implementation with our RDMA Channel-based zero-copy design using
latency and bandwidth microbenchmarks. We see that the two
implementations perform comparably for small and large
messages. However, the CH3-based design outperforms the RDMA
Channel-based design for mid-sized messages (32K to 256K Bytes) in
bandwidth.

\begin{figure}[htbp]
\includegraphics[width=\columnwidth]{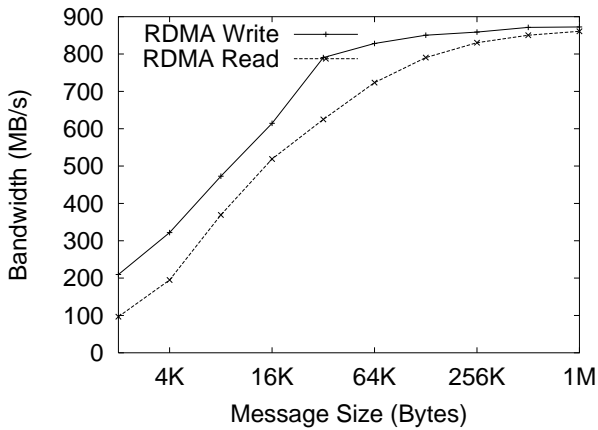}
\caption{InfiniBand Bandwidth}
\label{fig:vapi_readwrite}
\end{figure}

Figure~\ref{fig:vapi_readwrite} shows the bandwidth of RDMA read and
RDMA write at the InfiniBand VAPI level.  (VAPI is the programming
interface for our InfiniBand cards.)  With the current VAPI
implementation, RDMA write operations have a clear advantage over RDMA
read for mid-sized messages. Therefore, the fact that CH3-based design
outperforms RDMA Channel-based design for mid-sized messages is more
the result of the raw performance difference between RDMA write and
RDMA read than the designs themselves.

\section{Application-Level Evaluation}
\label{sec:perf}

In this section, we carry out an application-level evaluation of our
MPICH2 designs using NAS Parallel Benchmarks~\cite{NAS_BENCH}.  We run
class A benchmarks on 4 nodes and class B benchmarks on 8 nodes.
Benchmarks SP and BT require a square number of nodes. Therefore,
their results are only shown for 4 nodes.

\begin{figure}[htbp]
\includegraphics[width=\columnwidth]{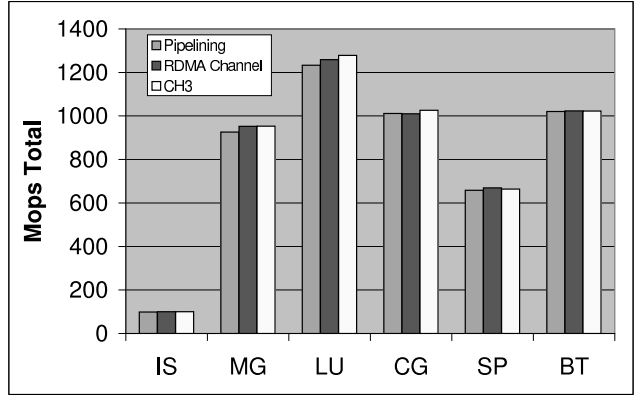}
\caption{NAS Class A on 4 Nodes}
\label{fig:mpich2_nas4}
\end{figure}

\begin{figure}[htbp]
\includegraphics[width=\columnwidth]{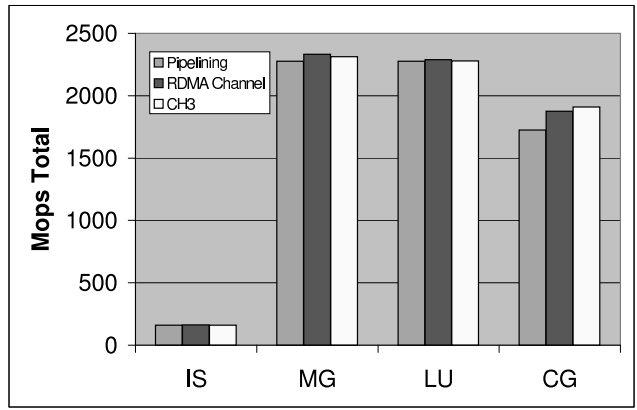}
\caption{NAS Class B on 8 Nodes}
\label{fig:mpich2_nas8}
\end{figure}

The results are shown in Figures~\ref{fig:mpich2_nas4} and
\ref{fig:mpich2_nas8}.  We have evaluated three designs: RDMA Channel
implementation with pipelining for large messages (Pipelining), RDMA
Channel implementation with zero-copy for large messages (RDMA
Channel), and CH3 implementation with zero-copy (CH3).  Although the
performance difference of these three designs is not much, we observe
that the pipelining design performs the worst in all cases. The RDMA
Channel-based zero-copy design performs very close to the the
CH3-based zero-copy design. On average, the CH3-based design performs
less than 1\% better on both 4 nodes and 8 nodes.

\section{Related Work}
\label{sec:related}

As the predecessor of MPICH2 and one of the most popular MPI
implementations, MPICH supports a similar implementation structure as
MPICH2. MPICH provides ADI2 (the second generation of Abstract Device
Interface) and Channel interface. Various implementations exist based
on these interfaces~\cite{MPI-MAD,MPIMyrinet,MPIQuadrics,MVICH}.  Our
MVAPICH implementation~\cite{MVAPICH_RDMA}, which exploits RDMA write
in InfiniBand, is based on the ADI2 interface.

Since MPICH2 is relatively new, there exists very little work
describing its implementations on different architectures. In
\cite{CH3}, a CH3-level implementation based on TCP/IP is
described. Work in \cite{MPICH2_DE} presents an implementation MPICH2
over InfiniBand, also using the CH3 interface.  However, in our paper,
our focus is on the RDMA Channel interface instead of the CH3
interface.  MPICH2 is designed to support both MPI-1 and MPI-2
standards.  There have been studies about supporting the MPI-2
standard, especially one-sided communication
operations~\cite{MPI2-SUN,MPI2-NEC}.  To date, we have concentrated
on supporting MPI-1 functions in MPICH2. We plan to explore the
support of MPI-2 functions in the future.

Because of its high bandwidth and low latency, the InfiniBand
Architecture has been used as the communication subsystem in a number
of systems other than MPI, such as distributed shared-memory systems
and parallel file systems~\cite{IB-PVFS,IB-DSM}.

The RDMA Channel interface presents a stream-based abstraction
somewhat similar to the traditional socket interface.  There have been
studies of how to implement user-level socket interface efficiently
over high-speed interconnects such as Myrinet, VIA, and Gigabit
Ethernet~\cite{SOCKET-GM,SOCKET-VIA,SOCKET-EMP}.  Recently, Socket
Direct Protocol (SDP)~\cite{SDP} has been proposed, which provides a
socket interface over InfiniBand.  The idea of our zero-copy scheme is
similar to the Z-Copy scheme in SDP.  However, there are also
differences between the RDMA Channel interface and the traditional
socket interface.  For example, {\em put} and {\em get} functions in
RDMA Channel interface are nonblocking, while functions in the
traditional sockets are usually blocking.  To support traditional
socket interface, one has to make sure the same semantics are
maintained.  We do not have to deal with this issue for the RDMA
Channel interface.

\section{Conclusions and Future Work}
\label{sec:conclusion}

In this paper, we present a study of using RDMA operations to
implement MPICH2 over InfiniBand. Our work takes advantage of the RDMA
Channel interface provided by MPICH2.

The RDMA Channel interface provides a very small set of functions to
encapsulate the underlying communication layer on which the whole
MPICH2 implementation is built. Consisting of only five functions, the
RDMA Channel interface is easy to implement for different
communication architectures. However, the question arises whether this
abstraction is powerful enough that one can still achieve good
performance.

Our study has shown that the RDMA Channel interface still provides the
implementors much flexibility. With optimizations such as
piggybacking, pipelining, and zero-copy, MPICH2 is able to deliver
good performance to the application layer. For example, one of our
designs achieves 7.6~$\mu$s latency and 857~MB/s peak bandwidth, which
come quite close to the raw performance of InfiniBand.  In our study,
we characterize the impact of each optimization by using latency and
bandwidth microbenchmarks. We have also conducted an application-level
evaluation using the NAS Parallel Benchmarks.

So far, our study has been restricted to a fairly small platform
consisting of 8 nodes.  In the future, we plan to use larger clusters
to study various aspects of our designs regarding scalability.
Another direction we are pursuing is to provide support for MPI-2
functionalities such as one-sided communication using RDMA and atomic
operations in InfiniBand.  We are also working on how to support
efficient collective communication on top of InfiniBand.

\bibliographystyle{latex8}
\bibliography{mydoc_bib}

\clearpage

\begin{figure*}

The submitted manuscript has been created in part by the University of
Chicago as Operator of Argonne National Laboratory ("Argonne") under
Contract No.  W-31-109-ENG-38 with the U.S. Department of Energy.  The
U.S. Government retains for itself, and others acting on its behalf, a
paid-up, nonexclusive, irrevocable worldwide license in said article
to reproduce, prepare derivative works, distribute copies to the
public, and perform publicly and display publicly, by or on behalf of
the Government.
\thispagestyle{empty}

\end{figure*}

\end{document}